\documentclass{aastex}
\usepackage{graphics}
\usepackage{epsfig}
\usepackage{emulateapj5}

\submitted{Accepted for Publication in \em{The Astrophysical Journal}}
\begin{document}

\title{Ages of A--type Vega--like stars from \emph{uvby\( \beta  \)} Photometry}

\author{Inseok Song, J.-P. Caillault,}

\affil{Department of Physics and Astronomy, University of Georgia, Athens GA 30602-2451
USA}

\email{song@physast.uga.edu, jpc@akbar.physast.uga.edu}

\author{David Barrado y Navascu\'{e}s,}

\affil{Departamento de F\'{i}sica Te\'{o}rica, Universidad Aut\'{o}noma de Madrid,
E--28049 Madrid SPAIN}

\email{barrado@pollux.ft.uam.es}

\author{John R. Stauffer}

\affil{Harvard Smithsonian Center for Astrophysics, 60 Garden St., Cambridge, MA 02138
USA}

\email{stauffer@amber.harvard.edu}

\shorttitle{Ages of A--type Vega--like stars from \emph{uvby\( \beta  \)} Photometry}

\shortauthors{Song et al.}

\slugcomment{submitted to The Astrophysical Journal}

\begin{abstract}
We have estimated the ages of a sample of A--type Vega--like stars by using
Str\"{o}mgren \emph{uvby\( \beta  \)} photometric data and theoretical evolutionary
tracks. We find that 13 percent of these A stars have been reported as Vega--like
stars in the literature and that the ages of this subset run the gamut from
very young (50~Myr) to old (1~Gyr), with no obvious age difference compared
to those of field A stars. We clearly show that the fractional IR luminosity
decreases with the ages of Vega--like stars.
\end{abstract}

\keywords{planetary systems --- circumstellar matter --- infrared: stars --- stars: early--type }

\section{Introduction}

There are several unusual sub--groups among the A--type stars, such as the metallic--line
stars (Am), the peculiar A stars (Ap), \( \lambda  \) Bootis type stars, and
shell stars \citep{AbtMorrell95}. Another class of stars with many members amongst
the A dwarfs is that of the Vega--like stars. Vega--like stars show excess IR
emission attributable to an optically thin dust disk around them. These disks
are believed to have very little or no gas \citep{LBA99}. It is very important
to know the ages and, hence, the evolutionary stages of these stars, since they
are believed to be signposts of exo--planetary systems or of on--going planet
formation. However, determining the ages of individual A--type stars is a very
difficult task. Some indirect age dating methods for A--type stars include the
use of late--type companions if any exist (HR~4796A and Fomalhaut; see \citealt{Stauffer95,Barrado97,myPhD})
or using stellar kinematic groups (Fomalhaut, Vega and \( \beta  \)~Pictoris;
see \citealt{David98} and \citealt{BSSC}). The use of Str\"{o}mgren \emph{uvby\( \beta  \)}
photometry \citep{ATF97}, however, provides a more direct and general determination
of the ages of A--type stars. 

The photometric \emph{uvby}\( \beta  \) system as defined by \citet{Stromgren63}
and \citet{CM66} allows for reasonably accurate determination of stellar parameters
like effective temperature \( T_{eff} \), surface gravity \( g \), and metallicity
for B, A, and F stars \citep[and references therein]{crawford79,NSW93}. The
\( T_{eff} \) and \( g \) values can then be used to estimate directly the
ages of stars when they are coupled with theoretical evolutionary tracks (though
for individual stars these estimates have relatively large error bars). 

In this letter, we describe our application of this technique to a volume limited
sample of 200 A stars.

\section{Method}

\subsection{\protect\( T_{eff}\protect \) and \protect\( \log g\protect \) determination}

Extensive catalogues of \emph{uvby}\( \beta  \) data have been published by
\citet{HM80}, \citet{Olsen83}, and \citet{OP84}. We have used these catalogues
and WEBDA\footnote{%
Web version of BDA (Open clusters database, \citealt{BDA}) http://obswww.unige.ch/webda
} databases to find \emph{uvby\( \beta  \)} photometry data for our sample of
A--type stars.

Numerous calibration methods of effective temperature and surface gravity using
\emph{uvby\( \beta  \)} photometry have been published \citep[and references therein] {NSW93,SD95}.
\citet{MD85}, in particular, demonstrate that their calibration yields \( T_{eff} \)
and \( \log g \) to a 1~\( \sigma  \) accuracy of \( 260 \)~K and \( 0.10 \)~dex,
respectively. However, as pointed out by \citet{NSW93}, \( \log g \) from \citeauthor{MD85}'s
calibration depends on the \( T_{eff} \) value while the most desirable calibration
method should not. Therefore, we used the \citet{MD85} grids with Napiwotzki
et al.'s gravity modification to eliminate the \( \log g \) dependence on \( T_{eff} \)
for early--type stars. The subsequent temperature calibration is in agreement
with the integrated--flux temperatures \( \left( T_{eff}=(\pi F/\sigma )^{1/4}\right)  \)
from \citet{Code}, \citet{Beeckmans}, and \citet{Malagnini} at the 1\% level
and the accuracy of \( \log g \) ranges from \( \approx 0.10 \) dex for early
A stars to \( \approx 0.25 \) dex for hot B stars \citep{NSW93}.

A rapidly rotating star has a surface gravity smaller at the equator than at
the poles and both the local effective temperature and surface brightness are
therefore lower at the equator than at the poles. Thus, in comparing a rotating
star with a non--rotating star of the same mass, the former is always cooler.
But the apparent luminosity change of a rotating star depends on the inclination
angle (\( i \)) such that a pole--on \( \left( i=0^{\circ }\right)  \) star
is brighter and an edge--on \( \left( i=90^{\circ }\right)  \) star is dimmer
than a non--rotating star \citep{Kraft}. In all cases, the combination of the
luminosity and temperature changes result in an older inferred age compared
to the non--rotating case. This effect is prominent in spectral types B and
A in which most stars are rapidly rotating (\( v\sin i\geq 100km/sec \)). Recently,
\citet{FB98} simulated the effect of stellar rotation on the Str\"{o}mgren
\emph{uvby\( \beta  \)} photometric indices. They concluded that the effect
of stellar rotation is to enhance the stellar main sequence age by an average
of \( 40\% \). Therefore, we included the stellar rotation correction suggested
by \citet{FB98}. However, their rotation correction schemes are available only
for stars with spectral type between approximately B7--A4. We extended the range
of rotation correction such that for stars earlier than B7, we used the correction
scheme for B7 stars and for stars later than A4, we used the correction scheme
for A4 stars. Therefore, stars earlier or later than the Figueras \& Blasi's
(1998) range will have more uncertain ages. 

Large uncertainties in estimated ages are mainly due to the large error in \( \log g \).
However, using a rotation correction scheme based on the projected stellar rotational
velocities \( \left( v\sin i\right)  \) rather than a scheme based on the true
stellar rotational velocities \( \left( v\right)  \) may have resulted in uncertainties
also. The stellar rotation decreases the effective temperature depending on
the inclination angle (small change of \( T_{eff} \) for \( i\approx 0^{\circ } \)
but large change of \( T_{eff} \) for \( i\approx 90^{\circ } \)) but the
current rotation correction scheme cannot distinguish between the case of large
\( v \) with small \( i \) and the case of small \( v \) with large \( i \).
Thus, rotation correction using \( v\sin i \) instead of \( v \) may cause
uncertainty in stellar ages.

\subsection{Ages of Open Clusters}

The theoretical evolutionary grids of \citet{Schaller92} were used to estimate
ages of stars from \( T_{eff} \) and \( \log g \). To verify that our age
dating method is working, we applied the method to a few open clusters with
ages determined by other methods -- \( \alpha  \)~Perseus (80~Myr), Pleiades
(125~Myr), NGC~6475 (220~Myr), M34 (225~Myr), and Hyades (660~Myr). The
ages for the first two clusters are based on recent application of the lithium
depletion boundary method (LDBM) (\citealt{PerAge} and \citealt{Stauffer99}
for \( \alpha  \)~Perseus; \citealt{Stauffer98} for Pleiades). The ages for
the other clusters are from upper main sequence isochrone fitting (UMSIF), and
are taken from \citet{JonesProsser} or \citet{Lynga}. The age scales based on
the two different methods (LDBM and UMSIF) are not yet consistent with each
other, and both have possible systematic errors. The current best UMS isochrone
ages for \( \alpha  \)~Perseus and the Pleiades are in the range 50--80 Myr
and 80--150 Myr.

\begin{figure}
\plotone{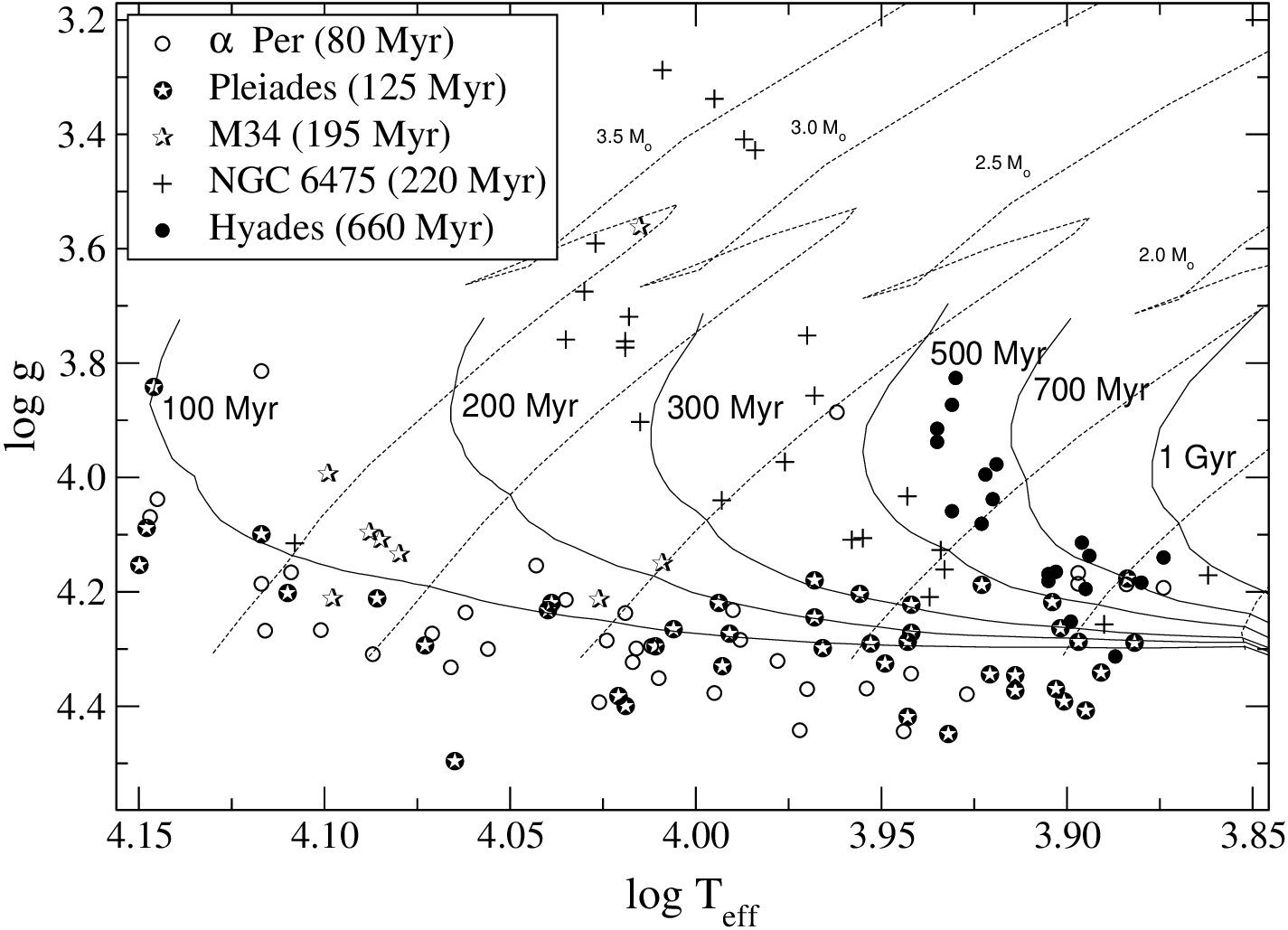}
\caption{Str\"{o}mgren \emph{uvby\protect\( \beta \protect \)} photometric age determination
of a few open clusters.\label{OpenCluster}}
\end{figure}

In Figure~\ref{OpenCluster}, one can see that the isochrones of these open
clusters are fairly well reproduced. However, there are some deviations from
the expected values. Stars that are younger than or close to 100~Myr, like
stars in \( \alpha  \) Perseus, tend to locate below the theoretical 100~Myr
isochrone. So we assigned an age of 50 Myr for the stars below the 100~Myr
isochrone. At intermediate ages, the open cluster data provide a mixed message
-- the M34 Strömgren age appears to be younger than the UMSIF age, whereas the
NGC~6475 Strömgren age seems older than the UMSIF age. This could be indicative
of the inhomogeneous nature of the ages (some from LDBM, some from relatively
old UMS models, some from newer models) to which we are comparing the Strömgren
ages. 

\begin{figure}
\plotone{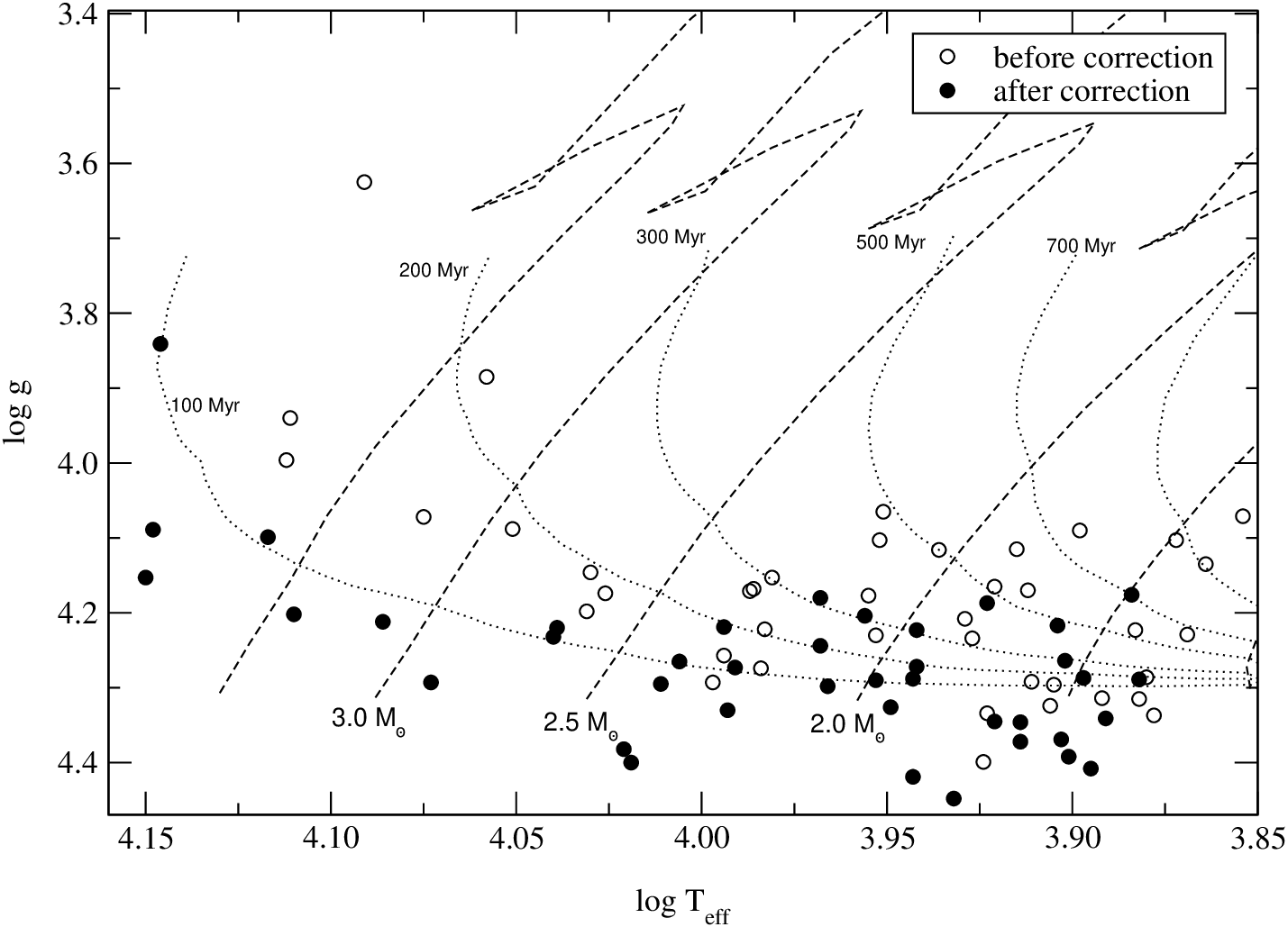}
\caption{Effect of stellar rotation correction applied to Pleiades stars.\label{roteffect}}
\end{figure}

If we could use \( v \) data instead of \( v\sin i \) and if one could make
a rotation correction scheme by using \( v \) values, then the new correction
scheme would tighten more stars for a given cluster to the locus of the cluster
compared to the uncorrected case. However, the \( v\sin i \) rotation correction
scheme used in this study shifts the loci of clusters and only moderately reduces
the standard deviations of ages (see, e.g., the case for the Pleiades in Figure~\ref{roteffect}).

\section{Field A stars and Vega--like stars}

We have identified 200 A dwarfs within 50 pc with known \( v\sin i \) values
and measured \emph{uvby\( \beta  \)} photometric indices. The distance limit
of 50~pc was chosen so that the photospheres of most A--type stars within the
given volume should be detected in the 12~\( \mu  \)m IRAS band and that the
volume should contain enough A--type stars to draw a statistically significant
result. Since rotation greatly affects the estimated stellar ages, we only included
stars with known \( v\sin i \) values (from SIMBAD) throughout this study.
\( T_{eff} \) and \( \log g \) values were calculated and corrected to account
for the rotation effects as described in the previous section. Among these A--stars,
26 have been identified as possible Vega--like stars by cross--indexing the
current list with Song's \citeyearpar{myPhD} master list of {}``proposed{}''
Vega--like stars. Estimated ages, along with other data --- spectral type, fractional
IR luminosity \( f \), \emph{uvby\( \beta  \)} photometric \emph{}data\emph{,}
and \( v\sin i \) --- are summarized in Table~\ref{AVegas}. The frequency
of Vega--like stars in our sample is 13\% in good agreement with the results
from other volume limited surveys: \( 14\pm 5\% \) from \citeauthor{Plets99}'s
\citeyearpar{Plets99} survey of the incidence of the Vega phenomenon among main
sequence and post main sequence stars and about \( 15 \)\% or more from the
review article on the Vega phenomenon by \citet {LBA99}. 

More than 95\% of our sample stars are listed in the IRAS Point Source Catalog
and/or Faint Source Catalog and were detected at least at 12~\( \mu  \)m,
and about 75\% of them were detected at 12 and 25~\( \mu  \)m. Based on the
12 and 25~\( \mu  \)m IRAS fluxes, we checked whether there could be more
IR excess stars besides the 26 already reported in the literature. Photospheric
IR fluxes at the IRAS bands were calculated by using

\begin{equation}
\label{eqn}
F_{\nu }=6.347\times 10^{4}\frac{\pi ^{2}R^{2}}{\lambda ^{3}}\frac{1}{\exp \left( \frac{14388}{\lambda T}\right) -1}\, [Jy]
\end{equation}
 where \( \pi  \) is parallax in arcseconds, \( R \) is stellar radius in
solar radii, \( \lambda  \) is wavelength in \( \mu  \)m, and \( T \) is
stellar effective temperature in Kelvins \citep{myPhD}. In Equation \ref{eqn},
\( R \) and \( T \) values were calculated from the \( M_{v} \) versus \( R \)
or \( T \) relations \citep{AQ} where \( M_{v} \) values were determined from
apparent visual magnitude (from SIMBAD) and \emph{Hipparcos} distance data.
Uncertainties of IR fluxes (\( \Delta F_{\nu } \)) were calculated from

\begin{equation}
\label{Ferror}
\Delta F_{\nu }=F_{\nu }\left( \pi _{\circ },R_{\circ },T_{\circ }\right) \left[ \frac{2\Delta \pi }{\pi }+\frac{2\Delta R}{R}\right] \, [Jy]
\end{equation}
where flux uncertainty due to \( \Delta \mathrm{T} \) is negligible (less than
0.02\% for a given 1\% error in T at 10,000K). Average flux uncertainties due
to \( \pi  \) and \( R \) uncertainties are 3\% and 4\%, respectively. If
we define the significance of IR excess (\( r_{\nu } \)) as excess IR flux
normalized by the uncertainty, then it can be calculated by \( r_{\nu }=(F_{IRAS}-F_{\nu })/\Delta F \)
where \( \Delta F \) is the total flux uncertainty due to \( \Delta F_{\nu } \)
and \( \Delta F_{IRAS} \) (\( F_{IRAS} \) and \( \Delta F_{IRAS} \) stand
for flux value and flux uncertainty value from the IRAS catalog, respectively).
\( \Delta F_{\nu } \) and \( \Delta F_{IRAS} \) were added in quadrature to
calculate the total flux uncertainty (\( \Delta F \)). We define the \emph{bona--fide}
Vega--like stars to be those that show significant IR excesses, \( r_{\nu }\geq 3.0 \),
at three or more IRAS bands, with the most prominent excess at 60~\( \mu  \)m.
We have found that 51 additional stars show significant IR excesses (\( r_{\nu }\geq 3.0 \))
at both 12 and 25~\( \mu  \)m. However, only 14 of them turned out to be legitimate
Vega--like star candidates. The other 37 stars are either luminosity class \( III \)
stars (whose IR excesses would not arise because of a circumstellar dust disk)
or stars whose excess radiation can easily be explained with a nearby companion
star within the IRAS beam. The new Vega--like candidates are summarized in Table~\ref{new}
with their \( r_{\nu } \) values at 12 and 25~\( \mu  \)m. Determining the
\( f \) values for the Vega--like candidates with only 12 and 25~\( \mu  \)m
IR flux measurements is difficult, because, for most of the cases, stellar photospheric
flux dominates compared to any excess at these wavelengths; thus a slight error
in the photospheric flux calculation results in a large error in \( f \) values.
For this reason, we have not taken these stars into account in our consideration
of \( f \) versus age relation (see below). 

The photospheric flux calculated from the \citeauthor{Plets99}'s \citeyearpar{Plets99} empirical
relation between the visual magnitude and the IRAS 12~\( \mu  \)m magnitude
is always higher than the photospheric flux values calculated by using equation~\ref{eqn};
thus the significance of the IR excess for most of the new Vega--like stars
falls below the 3~\( \sigma  \) threshhold when \citeauthor{Plets99}'s method
used. Therefore, these 14 new candidates have to be treated with care. We considered
two different sets of Vega--like stars: (1)~using \emph{al}l proposed Vega--like
stars (case A, N=26) and (2)~using only the \emph{bona--fide} stars (case B,
N=20). The second column of Table~\ref{AVegas} indicates the case(s) to which
the star belongs. Our conclusion, discussed below, does not depend on the choice
of case. 

We assume that all of the A stars in the sample are post--ZAMS stars. We make
that assumption because of simple timescale arguments (the ratio of \( <10 \)~Myr
old stars to the number of 100--300~Myr old stars should be of order \( <10/200 \)
or 5\%), and because we expect pre--ZAMS A stars to be located generally in
star forming regions, which would make them easy to identify. There is no obvious
age difference between field A--type stars and A--type Vega--like stars within
50pc with both groups running the gamut from very young (50~Myr) to old (1~Gyr).
This result (and those in \citet{Silverstone} and \citet{Song00}) contrasts
with \citeauthor{Habing99}'s (1999) claim of the Vega phenomenon ending sharply
at around 400~Myr.

We have checked whether a correlation exists between ages and dust properties
by comparing our estimated ages of A--type Vega--like stars and their fractional
IR luminosities, \( f\equiv (L_{IR}/L_{*}) \), found in \citet{myPhD}. Unfortunately,
a plot of \( f \) versus age is not very informative mainly because of the
large uncertainties of the estimated ages for individual stars. Therefore, we
divided the Vega--like stars into two groups, one for the stars younger than
200 Myr and the other for stars older than 200 My, and calculated each group's
average \( f \)--value (Table~\ref{fages}). Clearly, the younger A--type
Vega--like stars have higher \( f \) values compared to those of the older
ones (case--independent). However, we cannot more accurately quantify this relation
because the uncertainties in \( T_{eff} \) and \( \log g \) are large.

\section{Summary and Discussion}

In an attempt to determine the ages of A--type Vega--like stars, we have used
a technique involving \emph{uvby\( \beta  \)} photometry and theoretical \( \log T_{eff}-\log g \)
evolutionary tracks. In addition, we have applied corrections for the effects
of rapid rotation. As a test of this procedure, we have estimated the ages of
a few open clusters and find that our values are in good agreement with their
standard ages. We then applied this age dating method to the 200 A--type stars
within 50 pc with known \( v\sin i \) values. Thirteen percent of these A stars
have been reported as Vega--like stars in the literature and their ages run
the gamut of very young (50~Myr) to old (1~Gyr) with no obvious age difference
compared to the field A--stars. The younger Vega--like stars have higher \( f \)
values compared to those of the older ones.

Vega--like stars are closely related to the \( \lambda  \)~Boo stars. These
are metal--deficient A--stars with IR excesses. Vega itself is discussed as
a possible member of the \( \lambda  \)~Boo class \citep{HR}. An age determination
of \( \lambda  \)~Boo stars was presented by \citet{IB} based on the assumption
that \( \lambda  \)~Boo stars are main sequence stars. However, \citet{HR}
argue that \( \lambda  \)~Boo stars are probably pre--main sequence stars.
If \( \lambda  \)~Boo stars are indeed closely related to the Vega--like stars,
then, based on our determination of the main sequence nature of Vega--like stars,
it is likely that most of the \( \lambda  \)~Boo stars are main sequence stars.

\acknowledgements{We thank the referee, Ralf Napiwotzki, for his useful suggestions and for kindly
providing his FORTRAN code to extract \( T_{eff} \) and \( \log g \). IS and
JPC acknowledge the support of NASA through grant NAG5--6902. DBN has been partially
supported by spanish {\it Plan Nacional del Espacio}, under grant ESP98--1339-CO2.
We have used SIMBAD and Vizier databases.}


\begin{thebibliography}{36}
\expandafter\ifx\csname natexlab\endcsname\relax\def\natexlab#1{#1}\fi

\bibitem[\protect\astroncite{{Abt} \& {Morrell}}{1995}]{AbtMorrell95}
{Abt}, H.~A. \& {Morrell}, N.~I. 1995, {\em ApJS\/}, {\bf 99}, 135

\bibitem[\protect\astroncite{{Asiain} {\em et~al.\/}}{1997}]{ATF97}
{Asiain}, R., {Torra}, J., \& {Figueras}, F. 1997, {\em A\&A\/}, {\bf 322}, 147

\bibitem[\protect\astroncite{{Barrado y Navascués}}{1998}]{David98}
{Barrado y Navascués}, D. 1998, {\em \aap\/}, {\bf 339}, 831

\bibitem[\protect\astroncite{{Barrado y Navascués} {\em
  et~al.\/}}{1997}]{Barrado97}
{Barrado y Navascués}, D., {Stauffer}, J.~R., {Hartmann}, L., \&
  {Balachandran}, S.~C. 1997, {\em \apj\/}, {\bf 475}, 313

\bibitem[\protect\astroncite{{Barrado y Navascués} {\em et~al.\/}}{1999}]{BSSC}
{Barrado y Navascués}, D., {Stauffer}, J.~R., {Song}, I., \& {Caillault}, J.-P.
  1999, {\em \apjl\/}, {\bf 520}, L123

\bibitem[\protect\astroncite{{Basri} \& {Mart\'in}}{1999}]{PerAge}
{Basri}, G. \& {Mart\'in}, E.~L. 1999, {\em \apj\/}, {\bf 510}, 266

\bibitem[\protect\astroncite{{Beeckmans}}{1977}]{Beeckmans}
{Beeckmans}, F. 1977, {\em \aap\/}, {\bf 60}, 1

\bibitem[\protect\astroncite{{Code} {\em et~al.\/}}{1976}]{Code}
{Code}, A.~D., {Bless}, R.~C., {Davis}, J., \& {Brown}, R.~H. 1976, {\em
  \apj\/}, {\bf 203}, 417

\bibitem[\protect\astroncite{{Cox}}{2000}]{AQ}
{Cox}, N. 2000, {\em Allen's Astrophysical Quantities\/}, AIP Press,
  Springer-Verlag: New York, 4th ed.

\bibitem[\protect\astroncite{{Crawford}}{1979}]{crawford79}
{Crawford}, D.~L. 1979, {\em ApJ\/}, {\bf 84}, 12

\bibitem[\protect\astroncite{{Crawford} \& {Mander}}{1966}]{CM66}
{Crawford}, D.~L. \& {Mander}, J. 1966, {\em AJ\/}, {\bf 71}, 114

\bibitem[\protect\astroncite{{Figueras} \& {Blasi}}{1998}]{FB98}
{Figueras}, F. \& {Blasi}, F. 1998, {\em \aap\/}, {\bf 329}, 957

\bibitem[\protect\astroncite{{Habing} {\em et~al.\/}}{1999}]{Habing99}
{Habing}, H.~J., {\em et~al.\/} 1999, {\em \nat\/}, {\bf 401}, 456

\bibitem[\protect\astroncite{{Hauck} \& {Mermilliod}}{1980}]{HM80}
{Hauck}, B. \& {Mermilliod}, M. 1980, {\em A\&AS\/}, {\bf 40}, 1

\bibitem[\protect\astroncite{{Holweger} \& {Rentzsch-Holm}}{1995}]{HR}
{Holweger}, H. \& {Rentzsch-Holm}, I. 1995, {\em \aap\/}, {\bf 303}, 819

\bibitem[\protect\astroncite{{Iliev} \& {Barzova}}{1995}]{IB}
{Iliev}, I.~K. \& {Barzova}, I.~S. 1995, {\em \aap\/}, {\bf 302}, 735

\bibitem[\protect\astroncite{{Jones} \& {Prosser}}{1996}]{JonesProsser}
{Jones}, B.~F. \& {Prosser}, C.~F. 1996, {\em \aj\/}, {\bf 111}, 1193

\bibitem[\protect\astroncite{{Kraft}}{1970}]{Kraft}
{Kraft}, R.~P. 1970, in {\em Spectroscpic Astrophysics\/}, edited by
  C.~{Herbig}, University of California Press (Berkeley),  383--423

\bibitem[\protect\astroncite{{Lagrange} {\em et~al.\/}}{2000}]{LBA99}
{Lagrange}, A.~M., {Backman}, D.~E., \& {Artymowicz}, P. 2000, in {\em
  Protostars and planets {IV}\/}, edited by V.~{Mannings}, A.~P. {Boss}, \&
  S.~S. {Russell}, (Tucson: University of Arizona Press),  639--672

\bibitem[\protect\astroncite{{Lyng\aa}}{1987}]{Lynga}
{Lyng\aa}, G. 1987, Catalogue of Open Cluster Data 5th Ed., catalog No. VII/92A

\bibitem[\protect\astroncite{{Malagnini} {\em et~al.\/}}{1986}]{Malagnini}
{Malagnini}, M.~L., {Morossi}, C., {Rossi}, L., \& {Kurucz}, R.~L. 1986, {\em
  \aap\/}, {\bf 162}, 140

\bibitem[\protect\astroncite{{Mermilliod}}{1995}]{BDA}
{Mermilliod}, J.~C. 1995, in {\em Information and On-Line Data in Astronomy\/},
  edited by D.~{Egret} \& M.~A. {Albrecht}, Kluwer Academic Press, Dordrecht,
  127--138

\bibitem[\protect\astroncite{{Moon} \& {Dworetsky}}{1985}]{MD85}
{Moon}, T.~T. \& {Dworetsky}, M.~M. 1985, {\em MNRAS\/}, {\bf 217}, 305

\bibitem[\protect\astroncite{{Napiwotzki} {\em et~al.\/}}{1993}]{NSW93}
{Napiwotzki}, R., {Sch\"onberner}, D., \& {Wenske}, V. 1993, {\em A\&A\/}, {\bf
  268}, 653

\bibitem[\protect\astroncite{{Olsen}}{1983}]{Olsen83}
{Olsen}, E.~H. 1983, {\em A\&AS\/}, {\bf 54}, 55

\bibitem[\protect\astroncite{{Olsen} \& {Perry}}{1984}]{OP84}
{Olsen}, E.~H. \& {Perry}, C.~L. 1984, {\em A\&AS\/}, {\bf 56}, 229

\bibitem[\protect\astroncite{{Plets} \& {Vynckier}}{1999}]{Plets99}
{Plets}, H. \& {Vynckier}, C. 1999, {\em \aap\/}, {\bf 343}, 496

\bibitem[\protect\astroncite{{Schaller} {\em et~al.\/}}{1992}]{Schaller92}
{Schaller}, G., {Schaerer}, D., {Meynet}, G., \& {Maeder}, A. 1992, {\em
  \aaps\/}, {\bf 96}, 269

\bibitem[\protect\astroncite{{Silverstone}}{2000}]{Silverstone}
{Silverstone}, M.~D. 2000, Ph.D. thesis, University of California Los Angeles

\bibitem[\protect\astroncite{{Smalley} \& {Dworetsky}}{1995}]{SD95}
{Smalley}, B. \& {Dworetsky}, M.~M. 1995, {\em A\&A\/}, {\bf 293}, 446

\bibitem[\protect\astroncite{{Song}}{2000}]{myPhD}
{Song}, I. 2000, Ph.D. thesis, University of Georgia

\bibitem[\protect\astroncite{Song {\em et~al.\/}}{2000}]{Song00}
Song, I., Caillault, J.-P., {Barrado y Navascu\'es}, D., \& Stauffer, J.~R.
  2000, {\em ApJL\/}, {\bf 532}, 41

\bibitem[\protect\astroncite{{Stauffer} {\em et~al.\/}}{1995}]{Stauffer95}
{Stauffer}, J.~R., {Hartmann}, L.~W., \& {Barrado y Navascu\'es}, D. 1995, {\em
  \apj\/}, {\bf 454}, 910

\bibitem[\protect\astroncite{{Stauffer} {\em et~al.\/}}{1998}]{Stauffer98}
{Stauffer}, J.~R., {Schultz}, G., \& {Kirkpatrick}, J.~D. 1998, {\em \apjl\/},
  {\bf 499}, L199

\bibitem[\protect\astroncite{{Stauffer} {\em et~al.\/}}{1999}]{Stauffer99}
{Stauffer}, J.~R., {\em et~al.\/} 1999, {\em {\apj}\/}, {\bf 527}, 219

\bibitem[\protect\astroncite{{Str\"omgren}}{1963}]{Stromgren63}
{Str\"omgren}, B. 1963, {\em QJRAS\/}, {\bf 4}, 8

\end{thebibliography}

\onecolumn

{
\begin{table}

\caption{A--stars with IR excesses\label{AVegas}}
{\raggedright \begin{tabular}{ccccrcccccc|ccc}
\hline 
{\scriptsize HD}&
{\scriptsize Case}&
{\scriptsize Sp.}&
{\scriptsize \( f\equiv L_{IR}/L_{*} \)}&
\multicolumn{4}{c}{\emph{\scriptsize uvby\( \beta  \)} {\scriptsize photometric data}}&
{\scriptsize \( v\sin i \)}&
\multicolumn{2}{c|}{{\scriptsize corrected }}&
\multicolumn{3}{c}{{\scriptsize age (Myr)}}\\
\cline{5-8} \cline{10-11} \cline{12-14} 
{\scriptsize number}&
{\scriptsize }&
{\scriptsize type}&
{\scriptsize \( \times 10^{3} \)}&
\multicolumn{1}{c}{{\scriptsize \( b-y \)}}&
{\scriptsize \( m_{1} \)}&
{\scriptsize \( c_{1} \)}&
{\scriptsize \( \beta  \)}&
{\scriptsize (km/s)}&
{\scriptsize \( \log T_{e} \)}&
{\scriptsize \( \log g \)}&
{\scriptsize lower}&
{\scriptsize best}&
{\scriptsize upper}\\
\hline 
 {\scriptsize  3003}&
{\scriptsize A}&
{\scriptsize A0V}&
{\scriptsize 15}&
{\scriptsize 0.014}&
{\scriptsize 0.179}&
{\scriptsize 0.991}&
{\scriptsize 2.910}&
{\scriptsize 115}&
{\scriptsize 3.993}&
{\scriptsize 4.347}&
{\scriptsize --}&
{\scriptsize \( 50 \)}&
{\scriptsize \( 247 \)}\\
{\scriptsize 14055}&
{\scriptsize AB}&
{\scriptsize A1Vnn}&
{\scriptsize 0.048}&
{\scriptsize 0.005}&
{\scriptsize 0.166}&
{\scriptsize 1.048}&
{\scriptsize 2.889}&
{\scriptsize 240}&
{\scriptsize 4.028}&
{\scriptsize 4.188}&
{\scriptsize \( 50 \)}&
{\scriptsize \( 163 \)}&
{\scriptsize \( 245 \)}\\
{\scriptsize 38678}&
{\scriptsize AB}&
{\scriptsize A2Vann}&
{\scriptsize 0.17}&
{\scriptsize 0.054}&
{\scriptsize 0.188}&
{\scriptsize 0.996}&
{\scriptsize 2.877}&
{\scriptsize 230}&
{\scriptsize 3.990}&
{\scriptsize 4.189}&
{\scriptsize \( 50 \)}&
{\scriptsize \( 231 \)}&
{\scriptsize \( 347 \)}\\
{\scriptsize 39014}&
{\scriptsize AB}&
{\scriptsize A7V}&
{\scriptsize 0.11}&
{\scriptsize 0.126}&
{\scriptsize 0.182}&
{\scriptsize 0.961}&
{\scriptsize 2.790}&
{\scriptsize 225}&
{\scriptsize 3.937}&
{\scriptsize 3.797}&
{\scriptsize \( 522 \)}&
{\scriptsize \( 541 \)}&
{\scriptsize \( 663 \)}\\
 {\scriptsize 39060}&
{\scriptsize AB}&
{\scriptsize A3V}&
{\scriptsize 3}&
{\scriptsize 0.094}&
{\scriptsize 0.196}&
{\scriptsize 0.891}&
{\scriptsize 2.859}&
{\scriptsize 140}&
{\scriptsize 3.955}&
{\scriptsize 4.352}&
{\scriptsize --}&
{\scriptsize \( 50 \)}&
{\scriptsize \( 299 \)}\\
{\scriptsize 40932}&
{\scriptsize AB}&
{\scriptsize Am}&
{\scriptsize 0.23}&
{\scriptsize 0.093}&
{\scriptsize 0.200}&
{\scriptsize 0.981}&
{\scriptsize 2.853}&
{\scriptsize 20}&
{\scriptsize 3.919}&
{\scriptsize 3.966}&
{\scriptsize \( 565 \)}&
{\scriptsize \( 693 \)}&
{\scriptsize \( 693 \)}\\
{\scriptsize 50241}&
{\scriptsize AB}&
{\scriptsize A7IV}&
{\scriptsize 1.1}&
{\scriptsize 0.126}&
{\scriptsize 0.175}&
{\scriptsize 0.998}&
{\scriptsize 2.788}&
{\scriptsize 230}&
{\scriptsize 3.938}&
{\scriptsize 3.686}&
{\scriptsize \( 501 \)}&
{\scriptsize \( 664 \)}&
{\scriptsize \( 890 \)}\\
 {\scriptsize 71155}&
{\scriptsize AB}&
{\scriptsize A0V}&
{\scriptsize 0.062}&
{\scriptsize -0.007}&
{\scriptsize 0.158}&
{\scriptsize 1.026}&
{\scriptsize 2.896}&
{\scriptsize 130}&
{\scriptsize 4.013}&
{\scriptsize 4.205}&
{\scriptsize \( 50 \)}&
{\scriptsize \( 169 \)}&
{\scriptsize \( 266 \)}\\
 {\scriptsize 74956}&
{\scriptsize AB}&
{\scriptsize A1V}&
{\scriptsize 0.22}&
{\scriptsize 0.034}&
{\scriptsize 0.151}&
{\scriptsize 1.087}&
{\scriptsize 2.876}&
{\scriptsize 85}&
{\scriptsize 3.979}&
{\scriptsize 3.857}&
{\scriptsize \( 372 \)}&
{\scriptsize \( 390 \)}&
{\scriptsize \( 403 \)}\\
{\scriptsize 78045}&
{\scriptsize AB}&
{\scriptsize Am}&
{\scriptsize 0.03}&
{\scriptsize 0.077}&
{\scriptsize 0.188}&
{\scriptsize 0.960}&
{\scriptsize 2.871}&
{\scriptsize 40}&
{\scriptsize 3.928}&
{\scriptsize 4.184}&
{\scriptsize \( 50 \)}&
{\scriptsize \( 427 \)}&
{\scriptsize \( 610 \)}\\
{\scriptsize 91312}&
{\scriptsize AB}&
{\scriptsize A7IV}&
{\scriptsize 0.093}&
{\scriptsize 0.121}&
{\scriptsize 0.208}&
{\scriptsize 0.850}&
{\scriptsize 2.821}&
{\scriptsize 135}&
{\scriptsize 3.922}&
{\scriptsize 4.191}&
{\scriptsize \( 50 \)}&
{\scriptsize \( 414 \)}&
{\scriptsize \( 647 \)}\\
 {\scriptsize 95418}&
{\scriptsize AB}&
{\scriptsize A1V}&
{\scriptsize 0.0062}&
{\scriptsize -0.006}&
{\scriptsize 0.158}&
{\scriptsize 1.088}&
{\scriptsize 2.880}&
{\scriptsize 40}&
{\scriptsize 3.991}&
{\scriptsize 3.883}&
{\scriptsize \( 335 \)}&
{\scriptsize \( 358 \)}&
{\scriptsize \( 369 \)}\\
 {\scriptsize 99211}&
{\scriptsize AB}&
{\scriptsize A0V}&
{\scriptsize 0.012}&
{\scriptsize 0.117}&
{\scriptsize 0.194}&
{\scriptsize 0.894}&
{\scriptsize 2.822}&
{\scriptsize 145}&
{\scriptsize 3.925}&
{\scriptsize 4.069}&
{\scriptsize \( 392 \)}&
{\scriptsize \( 600 \)}&
{\scriptsize \( 684 \)}\\
{\scriptsize 102647}&
{\scriptsize A}&
{\scriptsize A3V}&
{\scriptsize 0.012}&
{\scriptsize 0.043}&
{\scriptsize 0.211}&
{\scriptsize 0.973}&
{\scriptsize 2.899}&
{\scriptsize 120}&
{\scriptsize 3.958}&
{\scriptsize 4.299}&
{\scriptsize --}&
{\scriptsize \( 50 \)}&
{\scriptsize \( 331 \)}\\
{\scriptsize 125162}&
{\scriptsize AB}&
{\scriptsize A0sh}&
{\scriptsize 0.042}&
{\scriptsize 0.051}&
{\scriptsize 0.183}&
{\scriptsize 0.999}&
{\scriptsize 2.894}&
{\scriptsize 100}&
{\scriptsize 3.966}&
{\scriptsize 4.188}&
{\scriptsize \( 50 \)}&
{\scriptsize \( 313 \)}&
{\scriptsize \( 451 \)}\\
{\scriptsize 135379}&
{\scriptsize A}&
{\scriptsize A3V}&
{\scriptsize 0.24}&
{\scriptsize 0.043}&
{\scriptsize 0.200}&
{\scriptsize 1.011}&
{\scriptsize 2.914}&
{\scriptsize 60}&
{\scriptsize 3.949}&
{\scriptsize 4.281}&
{\scriptsize \( 50 \)}&
{\scriptsize \( 166 \)}&
{\scriptsize \( 378 \)}\\
{\scriptsize 139006}&
{\scriptsize AB}&
{\scriptsize A0V}&
{\scriptsize 0.023}&
{\scriptsize -0.001}&
{\scriptsize 0.146}&
{\scriptsize 1.058}&
{\scriptsize 2.871}&
{\scriptsize 135}&
{\scriptsize 4.008}&
{\scriptsize 3.952}&
{\scriptsize \( 267 \)}&
{\scriptsize \( 314 \)}&
{\scriptsize \( 322 \)}\\
{\scriptsize 159492}&
{\scriptsize A}&
{\scriptsize A7V}&
{\scriptsize 0.094}&
{\scriptsize 0.102}&
{\scriptsize 0.204}&
{\scriptsize 0.883}&
{\scriptsize 2.858}&
{\scriptsize 80}&
{\scriptsize 3.927}&
{\scriptsize 4.322}&
{\scriptsize --}&
{\scriptsize \( 50 \)}&
{\scriptsize \( 419 \)}\\
{\scriptsize 161868}&
{\scriptsize AB}&
{\scriptsize A0V}&
{\scriptsize 0.068}&
{\scriptsize 0.015}&
{\scriptsize 0.173}&
{\scriptsize 1.051}&
{\scriptsize 2.898}&
{\scriptsize 220}&
{\scriptsize 4.011}&
{\scriptsize 4.199}&
{\scriptsize \( 50 \)}&
{\scriptsize \( 184 \)}&
{\scriptsize \( 277 \)}\\
{\scriptsize 172167}&
{\scriptsize AB}&
{\scriptsize A0V}&
{\scriptsize 0.013}&
{\scriptsize 0.003}&
{\scriptsize 0.157}&
{\scriptsize 1.088}&
{\scriptsize 2.903}&
{\scriptsize 15}&
{\scriptsize 3.987}&
{\scriptsize 4.031}&
{\scriptsize \( 267 \)}&
{\scriptsize \( 354 \)}&
{\scriptsize \( 383 \)}\\
{\scriptsize 172555}&
{\scriptsize AB}&
{\scriptsize A7V}&
{\scriptsize 0.9}&
{\scriptsize 0.112}&
{\scriptsize 0.200}&
{\scriptsize 0.839}&
{\scriptsize 2.839}&
{\scriptsize 175}&
{\scriptsize 3.942}&
{\scriptsize 4.376}&
{\scriptsize --}&
{\scriptsize --}&
{\scriptsize \( 50 \)}\\
{\scriptsize 178253}&
{\scriptsize A}&
{\scriptsize A0/A1V}&
{\scriptsize 0.064}&
{\scriptsize 0.018}&
{\scriptsize 0.184}&
{\scriptsize 1.060}&
{\scriptsize 2.889}&
{\scriptsize 225}&
{\scriptsize 4.008}&
{\scriptsize 4.112}&
{\scriptsize \( 164 \)}&
{\scriptsize \( 254 \)}&
{\scriptsize \( 316 \)}\\
{\scriptsize 181296}&
{\scriptsize AB}&
{\scriptsize A0Vn}&
{\scriptsize 0.14}&
{\scriptsize 0.000}&
{\scriptsize 0.157}&
{\scriptsize 1.002}&
{\scriptsize 2.916}&
{\scriptsize 420}&
{\scriptsize 4.133}&
{\scriptsize 4.898}&
{\scriptsize --}&
{\scriptsize --}&
{\scriptsize \( 50 \)}\\
{\scriptsize 192425}&
{\scriptsize A}&
{\scriptsize A2V}&
{\scriptsize 0.067}&
{\scriptsize 0.028}&
{\scriptsize 0.188}&
{\scriptsize 1.024}&
{\scriptsize 2.920}&
{\scriptsize 160}&
{\scriptsize 3.987}&
{\scriptsize 4.354}&
{\scriptsize --}&
{\scriptsize \( 50 \)}&
{\scriptsize \( 166 \)}\\
{\scriptsize 216956}&
{\scriptsize AB}&
{\scriptsize A3V}&
{\scriptsize 0.046}&
{\scriptsize 0.037}&
{\scriptsize 0.206}&
{\scriptsize 0.990}&
{\scriptsize 2.906}&
{\scriptsize 100}&
{\scriptsize 3.957}&
{\scriptsize 4.291}&
{\scriptsize \( 50 \)}&
{\scriptsize \( 156 \)}&
{\scriptsize \( 344 \)}\\
{\scriptsize 218396}&
{\scriptsize AB}&
{\scriptsize A5V}&
{\scriptsize 0.22}&
{\scriptsize 0.178}&
{\scriptsize 0.146}&
{\scriptsize 0.678}&
{\scriptsize 2.739}&
{\scriptsize 55}&
{\scriptsize 3.868}&
{\scriptsize 4.166}&
{\scriptsize \( 50 \)}&
{\scriptsize \( 732 \)}&
{\scriptsize \( 1128 \)}\\
\hline 
\end{tabular}\scriptsize \par}\end{table}
\par}

\begin{table}

\caption{New A--type Vega--like candidates\label{new}}
{\centering \begin{tabular}{cccccc}
\hline 
HD&
other&
Sp.&
\multicolumn{2}{c}{\( r_{\nu } \) }&
Remark\\
number&
name&
type&
12 \( \mu  \)m&
25 \( \mu  \)m&
\\
\hline 
2262&
\( \kappa  \) Phe&
A7V&
3.8&
3.3&
\\
6961&
33 Cas&
A7V&
3.7&
3.6&
\\
18978&
11 Eri&
A4V&
3.9&
3.5&
\\
20320&
13 Eri&
A5m&
4.1&
3.1&
SB\\
78209&
15 UMa&
A1m&
5.0&
3.7&
\\
87696&
21 LMi&
A7V&
3.4&
3.3&
\\
103287&
\( \gamma  \) UMa&
A0V&
4.8&
4.2&
SB\\
112185&
\( \epsilon  \) UMa&
A0p&
7.5&
6.4&
SB\\
123998&
\( \eta  \) Aps&
A2m&
5.0&
4.0&
\\
137898&
10 Ser&
A8IV&
4.0&
4.0\( ^{*} \)&
\\
141003&
\( \beta  \) Ser&
A2IV&
5.0&
4.0&
Double\\
192696&
33 Cyg&
A3IV--Vn&
5.7&
4.7&
SB\\
203280&
\( \alpha  \) Cep&
A7IV&
10.2&
5.5&
\\
214846&
\( \beta  \) Oct&
A9IV--V&
7.1&
5.8&
\\
\hline 
\end{tabular}\par}
{\par\centering \( ^{*} \)100~\( \mu  \)m excess, 25~\( \mu  \)m excess
\( r_{\nu }=1.9 \)\par}
\end{table}

\begin{table}

\caption{Ages and \protect\( f\protect \) values of Vega--like stars\label{fages}}
{\centering \begin{tabular}{c@{}l@{}lcc}
\hline 
\multicolumn{1}{c}{ Case}&
\multicolumn{2}{c}{Age }&
Number &
Average \( f \)\\
&
\multicolumn{2}{c}{range }&
of stars&
(\( \times 10^{3} \))\\
\hline 
A&
\( < \)&
200 Myr&
12&
1.79\\
&
\( > \)&
200 Myr&
14&
0.18\\
\hline 
B&
\( < \)&
200 Myr&
7&
0.71\\
&
\( > \)&
200 Myr&
13&
0.17\\
\hline 
\end{tabular}\par}\end{table}

\begin{figure}
\plotone{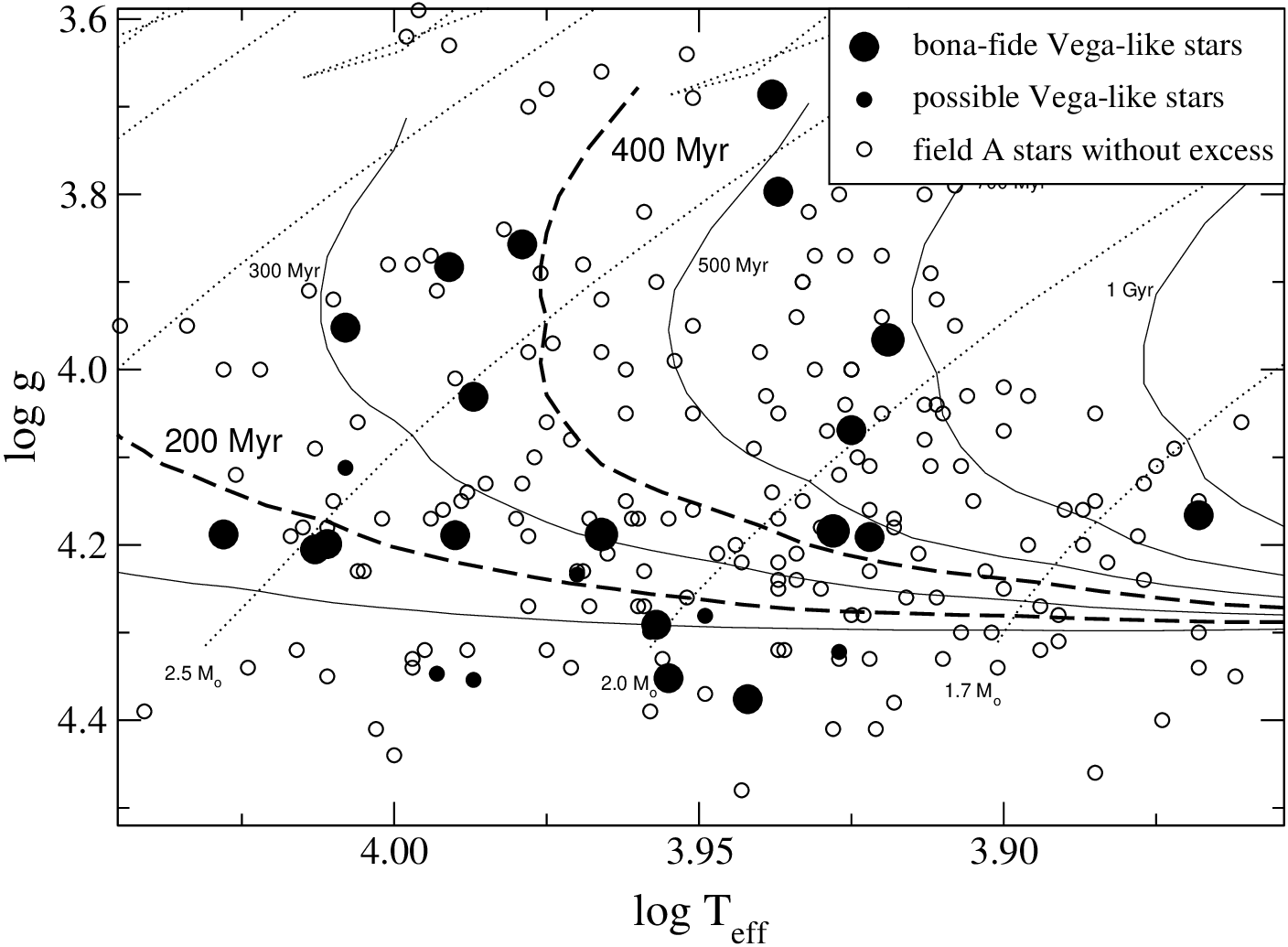}
\caption{Field A stars without excess and A--type Vega--like stars within 50 pc. Large
solid circles denote stars for case B only, small solid circles and large solid
circles form the set for case A. HD~181296 was not plotted because of its unusually
high \protect\( \log g\protect \) value.\label{Stromgren}}
\end{figure}

\end{document}